\documentclass[aps,prb,twocolumn,nowpacs,superscriptaddress,groupedaddress,amsmath,amssymb]{revtex4}
\usepackage {graphicx,epsfig,graphics,color}

\begin{document}



\title{Magnetic penetration-depth measurements of a suppressed superfluid density  of superconducting Ca$_{0.5}$Na$_{0.5}$Fe$_2$As$_2$ single crystals by proton irradiation}

\author{Jeehoon Kim}
\email[Corresponding author: ]{jeehoon@lanl.gov}
\affiliation{Los Alamos National Laboratory, Los Alamos, NM 87545}
\author{N. Haberkorn}
\affiliation{Los Alamos National Laboratory, Los Alamos, NM 87545}
\author{M. J. Graf}
\affiliation{Los Alamos National Laboratory, Los Alamos, NM 87545}
\author{I. Usov}
\affiliation{Los Alamos National Laboratory, Los Alamos, NM 87545}
\author{F. Ronning}
\affiliation{Los Alamos National Laboratory, Los Alamos, NM 87545}
\author{L. Civale}
\affiliation{Los Alamos National Laboratory, Los Alamos, NM 87545}
\author{E. Nazaretski}
\affiliation{Brookhaven National Laboratory, Upton, NY 11973}
\author{G. F. Chen}
\affiliation{Department of Physics, Renmin University of China, Beijing, 100872, China}
\author{W. Yu}
\affiliation{Department of Physics, Renmin University of China, Beijing, 100872, China}
\author{J. D. Thompson}
\affiliation{Los Alamos National Laboratory, Los Alamos, NM 87545}
\author{R. Movshovich}
\affiliation{Los Alamos National Laboratory, Los Alamos, NM 87545}

\date{\today}

\begin{abstract}

We report on the dramatic effect of random point defects,
produced by proton irradiation, on the superfluid density
$\rho_{s}$ in superconducting Ca$_{0.5}$Na$_{0.5}$Fe$_2$As$_2$
single crystals. The magnitude of the suppression is inferred
from measurements of the temperature-dependent magnetic
penetration depth $\lambda(T)$ using magnetic force microscopy.
Our findings indicate that a radiation dose of
2$\times$10$^{16}$cm$^{-2}$ produced by 3 MeV protons results in
a reduction of the superconducting critical temperature $T_{c}$ by
approximately 10$\%$.
In contrast, $\rho_{s}(0)$ is suppressed by approximately 60$\%$.
This break-down of the Abrikosov-Gorkov theory may be explained
by the so-called ``Swiss cheese model'', which accounts for the
spatial suppression of the order parameter near point defects
similar to holes in Swiss cheese. Both the slope of the upper
critical field and the penetration depth
$\lambda(T/T_{c})/\lambda(0)$ exhibit similar temperature
dependences before and after irradiation. This may be due to a
combination of the highly disordered nature of
Ca$_{0.5}$Na$_{0.5}$Fe$_2$As$_2$ with large intraband and
simultaneous interband scattering as well as the $s^\pm$-wave
nature of short coherence length superconductivity.

\end{abstract}

\maketitle

\section{Introduction}
Proximity of the superconducting and magnetic states in
iron-based superconductors has stimulated extensive studies of the
gap nature,\cite{Hashimoto,Ishida,Martin} order-parameter
symmetry,\cite{Tanatar,Kuroki,Nakayama} and the pairing
mechanisms in these materials.\cite{Mazin} The response of the
superconducting condensate to impurities is sensitive to the
symmetry of the superconducting state, and their influence has
been widely investigated to gain better understanding of the
nature of the order parameter in both low- and high-temperature
unconventional superconductors.
\cite{Anderson,Pethick,Annett,Nakajima,Kim}

The Abrikosov-Gor'kov (AG) theory\cite{Abrikosov} explains the
effects of impurities in the low-$T_{c}$ superconductors, where a
large superconducting coherence length $\xi$ effectively averages
the suppression of order parameter at the impurity sites over
many impurities, leading to a uniformly suppressed order
parameter. However, the AG theory breaks down when applied to the
effect of disorder on superconducting properties in the cuprates
superconductors,\cite{Annett} where $\xi$ is short and comparable
to the average spacing between disorder centers. The order
parameter is therefore suppressed locally at the impurity site
and has a chance to recover between impurities. The influence of
disorder on the superfluid density $\rho_{s}$ in cuprates is well
described by the so called ``Swiss cheese'' model, which
considers spatial dependence of the order parameter and its
strong suppression near
defects.\cite{Byers,Balatsky1,Flatte,Salkola,Hettler,Zhu,Balatsky2}
In iron-based systems, where superconductivity exhibits both
$s$-wave characteristics and a small coherence length, the
situation is between the low-temperature and high-temperature
superconductors. Consequently these systems pose an intriguing
question of how the effect of disorder on $T_{c}$ and the
superfluid density in these compounds compares to that in
conventional BCS superconductors and cuprates.\cite{Nakajima}

Recently, two irradiation experiments on Co-doped
BaFe$_{2}$As$_{2}$ (Co-122) were performed to study the influence
of disorder.\cite{Kim,Nakajima} The temperature-dependent
penetration depth measurements suggested an $s^{\pm}$ state, with
strong nonmagnetic scattering in the unitary limit,\cite{Kim}
whereas transport measurements showed an $s^{++}$ state with weak
scattering in the Born limit.\cite{Nakajima} Both experiments
showed a relatively small suppression of $T_{c}$ caused by
nonmagnetic impurities induced by irradiation; these findings are
consistent with an $s^{++}$ state, since superconductivity with a
sign changing order parameter is quite sensitive to nonmagnetic
impurities.\cite{Bang,Vorontsov,Gordon} Reports in several
iron-arsenide systems by different experimental techniques are
consistent with theoretical predictions of $s$ wave, potentially
nodal s-wave or sign reversing s-wave.\cite{Kuroki,Mazin}

In this work we investigate the influence of random point defects
introduced by proton irradiation on $\lambda(T)$ in
Ca$_{0.5}$Na$_{0.5}$Fe$_2$As$_2$ (CNFA) single crystals.  We use
the magnetic force microscopy (MFM) technique to determine
absolute values of $\lambda(T)$.\cite{Jeehoon PRB,Jeehoon,Jeehoon
MgB2,Xu1995,Coffey1995} The CNFA single crystals, showing homogeneity, have been grown
with a self-flux technique. Details of the sample preparation and
characterization can be found elsewhere.\cite{Haberkorn}

\section{Experiment}

The 3 MeV protons are known to produce between one and a few tens
of atomic displacements,\cite{Civale} creating random point
defects as well as nanoclusters with typical dimensions of few
nanometers. The CNFA sample was irradiated with the total proton
dose of 2$\times$10$^{16}$ cm$^{-2}$, which corresponds to an
average distance ($d$) between defects of  2.8 nm.\cite{Haberkorn
2012} The sample was cleaved, and its thickness measured to be
around 28 $\mu$m, which is smaller than the penetration range of
40 $\mu$m for the 3 MeV proton beam. Electrical resistivity in
both unirradiated and irradiated samples were measured using a
standard four-probe technique. The sample was mounted in a
rotatable probe and measurements were performed in magnetic
fields varying between 0 and 9 T. MFM measurements described here
were performed in a home-built low-temperature MFM
apparatus.\cite{Nazaretski RSI 2009} Three samples, CNFA,
irradiated CNFA (ICNFA), and a Nb reference film were loaded and
investigated in a comparative experiment within a single
cool-down. The magnetic stray field calibration was performed by
imaging vortices in a Nb reference as a function of applied
magnetic field.\cite{Jeehoon PRB} Measurements of $\lambda$ were
performed using the Meissner response technique.\cite{Jeehoon
PRB,Jeehoon,Jeehoon MgB2} The Meissner response curves were first
measured as a function of the tip-sample separation in the Nb
reference with known $\lambda(T)$. Subsequently, the cantilever
was moved to a sample of interest and the Meissner response
curves were acquired. Direct comparison of measured curves yields
the absolute value of $\lambda$ in a sample under investigation.
Details of experimental technique are described
elsewhere.\cite{Jeehoon PRB,Jeehoon,Jeehoon MgB2}

The reference Nb thin film ($T_{c}\approx$ 8.8 K) has a thickness
of 300 nm and was grown by electron beam deposition. The $T_{c}$
of CNFA from transport measurements is 19.4 K and that of ICNFA is
17.8 K. The width of the superconducting transition did not change
after irradiation. No upturn in resistivity was observed at low
temperatures, indicating that irradiation by protons results in
the formation of nonmagnetic point-like scattering
centers.\cite{Martin 2009} The MFM measurements were performed
using a high-resolution Nanosensors cantilever\cite {Nanosensors}
that was polarized along the tip axis in a 3 T magnetic field.
Both Nb and CNFA samples were zero-field cooled for Meissner
experiments; a magnetic field of a few Oe was applied above
$T_{c}$, followed by cooling for vortex imaging experiments.

\begin{figure}
\centering
\includegraphics [trim=0 0 0 2cm,clip=true,angle=0,width=8.5cm] {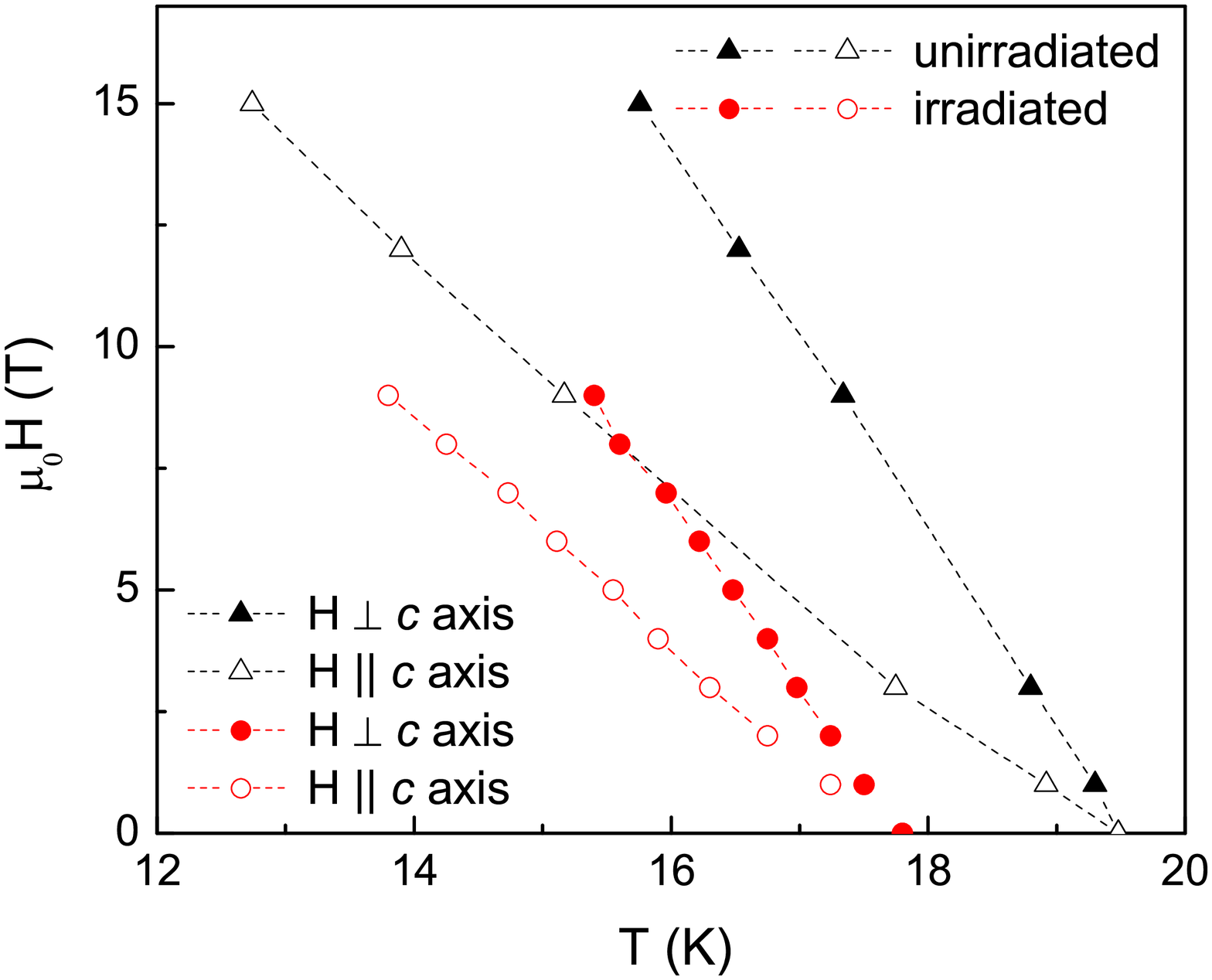}
\caption{\label{f:Hc2} (color online) Temperature dependent
H$_{c2}$ along the $c$ axis and within the $ab$ plane in
unirradiated Ca$_{0.5}$Na$_{0.5}$Fe$_2$As$_2$ (black triangles),
taken from Ref. 24, and proton irradiated
Ca$_{0.5}$Na$_{0.5}$Fe$_2$As$_2$ (red circles).}
\end{figure}

\section{Results}

\subsection{H$_{c2}(T)$ measurements}

Figure~\ref{f:Hc2} shows the upper critical field $H_{c2}(T)$
with ${\bf H}\parallel c$ and ${\bf H}\perp c$ (within the {\it
ab} plane) for CNFA and ICNFA. $H_{c2}(T)$ is linear in both
samples and for both directions, with average slopes of
$\beta^{ab} = -\frac{\partial H^{ab}_{c2}}{\partial T}\big
|_{T_{c}}=$ 4 T/K and $\beta^c = -\frac{\partial
H^{c}_{c2}}{\partial T}\big |_{T_{c}}=$ 2.2 T/K for CNFA, and
$\beta^{ab} = -\frac{\partial H^{ab}_{c2}}{\partial T}\big
|_{T_{c}}=$ 3.8 T/K and $\beta^c = -\frac{\partial
H^{c}_{c2}}{\partial T}\big |_{T_{c}}=$ 2.3 T/K for ICNFA. A
modest superconducting anisotropy parameter $\gamma =
{{\beta^{ab}}\over{\beta^c}} = 1.85-1.65$ for both CNFA and ICNFA
samples points toward a three-dimensional behavior. The
superconducting coherence length $\xi$ can be expressed in the
Ginzburg-Landau region as $\xi_{GL}(T) \approx
\xi_0/\sqrt{1-T/T_c}$. In the case of a one-band model or two
weakly coupled bands with similar  Fermi surface properties and
pairing interactions the zero-temperature in-plane coherence
length $\xi^{ab}_{0}$ and out-of-plane coherence length
$\xi^{c}_{0}$ are given by the slope of the upper critical
field:\cite{Haberkorn,Bauer2009}
$(\xi^{ab}_{0})^2\approx{\Phi_{0}}/{2\pi T_{c} \beta^{c}}$ and
$(\xi^{c}_{0})^2\approx{\Phi_{0}}/{2\pi T_{c} \beta^{ab}}$. We
obtain the zero-temperature Ginzburg-Landau values of
$\xi_{CNFA}^{ab}(0)$ = 2.8 nm and $\xi_{ICNFA}^{ab}(0)$ = 2.8 nm,
which are similar in magnitude to the short-coherence length
cuprate and PuCoGa$_5$ superconductors. Within our measurement
uncertainty no appreciable change of the coherence length took
place after irradiation, although $T_{c}$ is suppressed by 10\%.

\begin{figure}
\centering
\includegraphics [trim=0 0 0 9cm,clip=true,angle=0,width=9.0cm] {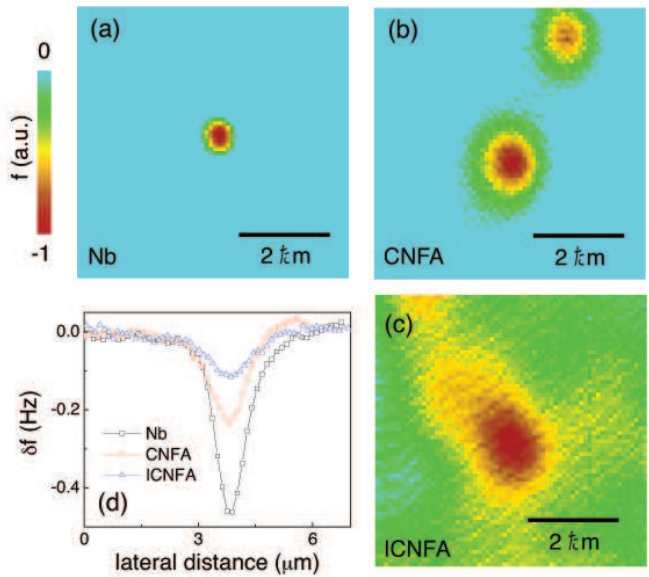}
\caption {\label{f:vortex} (color online) Single vortex images in
(a) the Nb reference, (b) the unirradiated
Ca$_{0.5}$Na$_{0.5}$Fe$_2$As$_2$, and (c) the irradiated
Ca$_{0.5}$Na$_{0.5}$Fe$_2$As$_2$. (d) Comparison of single vortex
profiles obtained from (a), (b), and (c). All images were
obtained under the same experimental conditions in a single-cool
down with the tip lift height of 300 nm at 4 K. The color scale
bar refers to (a)-(c).}
\end{figure}

\begin{figure}
\centering
\includegraphics [trim=1.5cm 0 0 10cm,clip=true,angle=0,width=8.5cm] {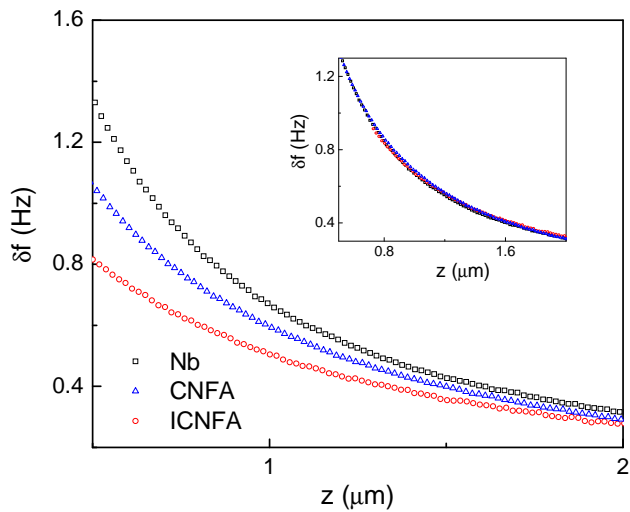}
\caption{\label{f:absolute} (color online) Meissner response
curves obtained from (a) the Nb reference (blue diamonds), (b)
the unirradiated Ca$_{0.5}$Na$_{0.5}$Fe$_2$As$_2$ (green
diamonds), and (c) the irradiated
Ca$_{0.5}$Na$_{0.5}$Fe$_2$As$_2$ (red diamonds) at 4 K. The
different slopes of the Meissner curves obtained from each sample
indicate a systematic change of $\lambda$. The inset: The
Meissner curves for the unirradiated and irradiated samples are
shifted along the horizontal axis to overlay the Meissner curve
for the reference Nb sample. The difference of the penetration
depths $\Delta\lambda$ can be obtained from the values of the
shift.}
\end{figure}

\subsection{$\lambda(T)$ measurements}

Prior to measurements of the absolute values of $\lambda(T)$,
vortex images were obtained under the same experimental
conditions for all samples. These measurements yield information
about homogeneity of CNFA and ICNFA samples on a submicron scale
($\sim 100$ nm). The well-formed single vortices in Nb and CNFA
suggest the homogeneity of the sample; however, the irregular
shape of single vortex in ICNFA (elongated vortex in the diagonal
direction of the image) suggests the presence of inhomogeneity in
the superfluid density on a sub-micron scale, which may be
related to impurities introduced from irradiation. We employed
the following imaging procedure: First, a single vortex in the Nb
sample was obtained at 4 K after the stray field calibration of
the MFM system.\cite{Jeehoon PRB} Second, the MFM tip was moved
on to CNFA and a single vortex image  obtained, and third, a
single vortex image was obtained after the tip was moved on to
ICNFA as shown in Figs.~\ref{f:vortex}(a), (b), and (c). The line
profile for each of the single vortices is shown in
Fig.~\ref{f:vortex}(d). The intensity of the vortex center in
different samples correlates with the magnitude of $\lambda$,
since all images were taken under the same conditions and with
the same tip. Lower intensity corresponds to a larger $\lambda$;
therefore, $\lambda$ in ICNFA is much larger than that in the Nb
reference. In addition, the magnitude of $\lambda$ among the
superconducting samples can be inferred from the relative size of
a single vortex: The larger the size, the larger is $\lambda$.
Therefore, $\lambda$ in ICNFA, showing the largest vortex size,
is the biggest among them.

\begin{figure}
\centering
\includegraphics [trim=0.5cm 0 0 10cm,clip=true,angle=0,width=9.0cm] {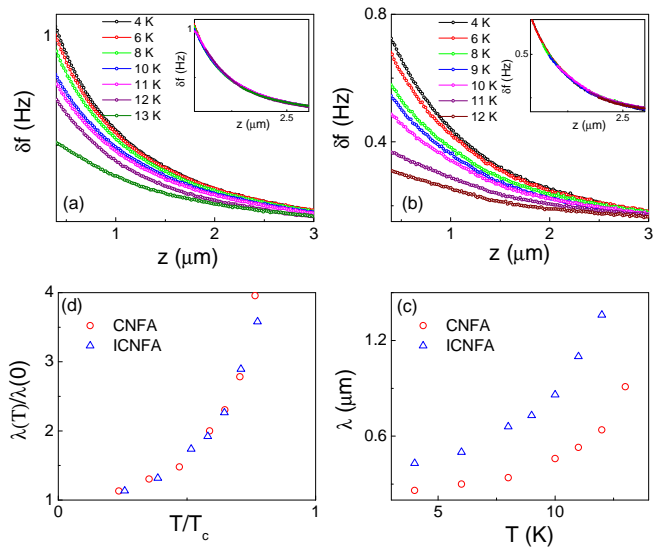}
\caption{\label{f:lambda-T} (color online) Temperature dependent
Meissner curves for (a) the unirradiated
Ca$_{0.5}$Na$_{0.5}$Fe$_2$As$_2$  and (b) the irradiated
Ca$_{0.5}$Na$_{0.5}$Fe$_2$As$_2$ samples. Insets in (a) and (b)
show overlaid temperature dependent Meissner curves at 4 K,
validating our procedure for extracting $\lambda(T)$. (c)
Temperature dependent $\lambda(T)$ in both unirradiated and
irradiated Ca$_{0.5}$Na$_{0.5}$Fe$_2$As$_2$ samples determined
from (a) and (b). (d) $\lambda(T)$ from (c) normalized by the
$T=0$ value as a function of the normalized temperature.}
\end{figure}

To extract absolute values of $\lambda$ in ICNFA we performed the
Meissner response measurements as described above. The Meissner
curves as a function of the tip-sample separation were obtained
in all three samples, Nb, CNFA, and ICNFA (see
Fig.~\ref{f:absolute}). The decay rate of the frequency shift
$\delta f$ as a function of the tip-sample separation $z$ provides
the relative magnitude of $\lambda$, {\it i.e.}, the higher the
rate $d(\delta f)/dz$ the larger the $\lambda$. In bulk and thick
films, the Meissner response force obeys a universal power-law
dependence with tip-to-sample distance.\cite{Xu1995,Coffey1995}
The force is given by $F_{Meissner}=A\times f(\lambda+z)$, where
$z$ is the tip-to-sample distance, $A$ is a pre-factor containing
information about the geometry of the magnetic tip, and $f(z)\sim
1/z^3$. By shifting the $f^{ICNFA}(z)$ data with respect to
distance in order to overlay it with the $f^{Nb}(z)$ curve, one
can obtain the absolute values of
$\lambda_{ICNFA}(T)=\lambda_{Nb}(T)+\Delta\lambda(T)$, where
$\Delta\lambda(T)$ is the magnitude of the shift. The shift
$\Delta\lambda$ between the Nb and ICNFA data equals 320 nm,
resulting in $\lambda_{ICNFA}(0)=\lambda_{Nb}(0)+\Delta\lambda(0)
= 110\;\text{nm} + 320\;\text{nm}=430$ nm. Using the same
procedure we also obtained
$\lambda_{CNFA}(0)=\lambda_{Nb}(0)+\Delta\lambda(0) =
110\;\text{nm}+150\;\text{nm} =260$ nm. Our experimental error is
around $10\%$ and depends on the magnitude of $\lambda$ and the
system noise level. A key result of this work is that the $\lambda(0)$ values before and after irradiation differ
significantly. This is in  stark contrast to both the coherence
length $\xi$, which shows little change after irradiation, as
well as the small suppression in $T_c$ of 10\%. The Meissner force MFM measurements of the ICNFA sample were performed after cleaving followed by irradiation. The ICNFA sample was remeasured after polishing. Both measurements showed the same $\lambda$ within experimental uncertainty. This indicates that irradiation does not noticeably affect sample quality. Therefore we can neglect the degradation of the sample surface for Meissner screening currents. It should be noted that our parameter free method of using the Nb reference sample is based on the assumption of a universal scaling function $F(z)$ for the Meissner force. This approach is valid for type-\MakeUppercase{\romannumeral 2} superconductors, where the electromagnetic response is local, {\it i.e.},  $\kappa = \lambda/\xi \gg 1$. Here we neglected higher order corrections in $1/\kappa$. Our Nb film has $\kappa = \lambda/\xi=110 nm/10 nm\approx 10$. The large $\kappa$ value in Nb allows direct comparison of Meissner responses between the Nb reference and CNFA, which results in good agreement by overlaying the Meissner curves, shown as insets in Fig.~\ref{f:absolute} and Figs.~\ref{f:lambda-T}(a)-(b). Our novel method of using a reference sample is justified a posteriori because the Meissner curves would not overlay with one another just by shifting them.

The temperature-dependent Meissner response curves measured in
both CNFA and ICNFA samples are shown in
Figs.~\ref{f:lambda-T}(a) and (b). The gradual variation of the
Meissner curves as a function of temperature indicates a
systematic change of $\lambda(T)$. The insets in (a) and (b) show
the Meissner curves obtained at different temperatures but shifted
to lie on top of the Meissner curve taken at $T=4$ K; the curves
overlay each other very well. The shift value for a given $T$ to
$T=4$ K along the horizontal axis allows one to calculate
$\lambda(T)$ at $T$. The resulting $\lambda(T)$ and normalized
$\lambda(T)/\lambda(0)$ in both samples are shown in
Figs.~\ref{f:lambda-T}(c) and (d), respectively. Results indicate
that $\lambda(T)$ increases after proton irradiation; however, the
dependence of $\lambda(T)/\lambda(0)$ on the normalized
temperature $T/T_c$ is the same for both samples within our
experimental uncertainty. The penetration depth exhibits the
typical power-law behavior $\Delta\lambda(T)/\lambda(0) \sim T^n$
with $n \approx 2$ reported previously for doped iron-arsenide
superconductors.\cite{Martin}

\section{Discussion}

\begin{figure}
\centering
\includegraphics [trim=0 0 0 0cm,clip=true,angle=0,width=7.5cm] {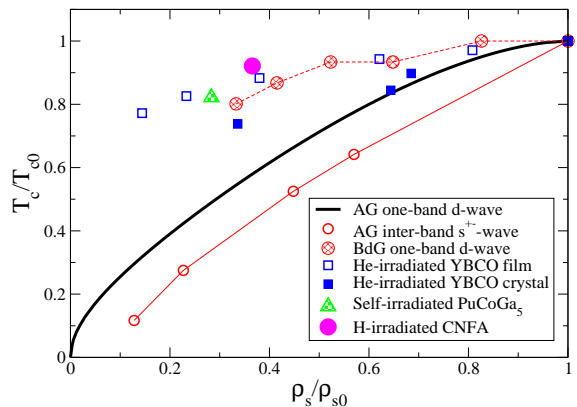}
\caption{\label{f:swiss} (online color) Uemura plot of the
superfluid density in disordered short coherence length
superconductors. $T_{c0}$ and $\rho_{s0}$ are values obtained
from a pristine crystal; $T_{c}$ and $\rho_{s}$ are those measured
after irradiation.  The solid circle represents the proton
irradiated CNFA obtained in this work. For comparison we plot
results of the one-band AG and BdG (Swiss cheese)
calculations\cite{Das} for $d$-wave pairing, two-band AG
$s^\pm$-wave calculations\cite{Vorontsov}, and experimental
results for self-irradiated PuCoGa$_5$ \cite{Ohishi2007} and
helium irradiated YBCO samples. \cite{Basov1994,Moffat1997}}
\end{figure}

The radiation dose of 2$\times$10$^{16}$ cm$^{-2}$ produced by 3
MeV protons in a Ca$_{0.5}$Na$_{0.5}$Fe$_2$As$_2$ sample causes
the suppression of the superfluid density $\rho_{s}(0)\approx
1/\lambda^{2}(0)$ by about 60$\%$ whereas $T_{c}$ is only suppressed by 10$\%$. We plot the value of the normalized
$\rho_s(0)$ for ICNFA as a solid circle in the Uemura
plot\cite{Uemura 1989}  of disordered superconductors in
Fig.~\ref{f:swiss}, as well as theoretical results of one-band AG
for d-wave pairing (solid line) and two-band AG calculations for
$s^{\pm}$ pairing (red open circles).\cite{Vorontsov} Also shown
are the BdG (Bogoliubov-de Gennes) calculations for $d$-wave
pairing (red hatched circles), a Swiss cheese model far from the
AG theory.~\cite{Das} Our result bears similarity to the data for
self-irradiated PuCoGa$_5$ \cite{Ohishi2007} and He-irradiated
YBCO high-temperature superconductor, showing that $T_{c}$ is
strongly immune to disorder relative to
$\rho_s(0)$,\cite{Basov1994,Moffat1997} contrary to the
conventional AG theory for $d$-wave paring. By analogy we argue
that the break-down of the AG theory is accounted for by the
Swiss cheese model within the BdG lattice theory of
short-coherence length superconductors,\cite{Das} which shows an
abrupt suppression of the order parameter near point defects.
This model describes the spatial dependence of the local density
of states and the order parameter in the vicinity (within a few
lattice constants) of a point-like nonmagnetic impurity in the
strong scattering limit, similar to holes  in Swiss cheese. Franz
and coworkers\cite{Franz} also reported the break down of the AG
theory and strong suppression of $\rho_{s}(0)$ for $d$-wave
paring. The effect is stronger in samples with small $\xi/a_{0}$
ratio ($a_{0}$ is the lattice constant). In the opposite limit,
the AG theory is valid and the order parameter is then suppressed
uniformly in the entire sample because $\xi \gg a_0, d $, with
$d$ the average distance between impurities. In our sample the
ratio of $\xi_{0}/a_{0}$ is approximately 7, $\xi\approx 2.8$ nm
and $d$ is about 2.8 nm, justifying the Swiss cheese scenario.

It is worth noting that the $T$ dependence of $\lambda(T)$ remains
the same after irradiation as shown in Fig.~\ref{f:lambda-T} (d),
while it changes in cuprates.\cite{Szotek,Prohammer1991,Kim1994}
This discrepancy may result from the nature of the multiband
$s$-wave paring as well as the highly disordered nature of CNFA
on the Ca/Na sites which lie above and below the iron layer.
The fact that the temperature behavior of $\lambda(T)$ is robust
after irradiation may be ascribed to large intraband scattering
with $s^\pm$ pairing and that the system itself is already in the
``dirty'' limit prior to irradiation, consistent with its short
coherence length and power-law dependence of $\lambda(T)$.
Additional disorder (mostly in the iron layer) by proton
irradiation therefore has little impact on the temperature
behavior of $\lambda(T)$, while added interband scattering is
detrimental to (increases) the absolute magnitude of $\lambda(0)$.

The pair-breaking effect due to nonmagnetic scattering in the AG
theory can be quantitatively analyzed using the normalized
scattering rate in conjunction with $\lambda$ given by:
$g^{\lambda}={\hbar\Delta\rho_{0}}/({2\pi
k_{B}T_{c0}\mu_{0}\lambda^{2}_{0}})$, where $\Delta\rho_{0}$ is
residual resistivity change induced by irradiation,
$\Delta\rho_{0}=\rho^{irr}_{0}-\rho^{unirr}_{0}$, $T_{c0}$ is the
critical temperature before irradiation, and $\lambda_{0}$ is the
penetration depth of the unirradiated sample.\cite{Nakajima} The
parameter $g^{\lambda}$ and $T_{c0}$ are expressed as $\text
{ln}(T_{c0}/T_{c})=\psi(1/2+g^\lambda
T_{c0}/(2T_{c}))-\psi(1/2)$, where $\psi(x)$ is the digamma
function, based on the $s^{\pm}$ scenario.\cite{Chubukov} This
pair-breaking result for $T_c$ is similar to that for conventional
$s$-wave with magnetic impurities or $d$-wave with nonmagnetic
impurities. Here the critical scattering rate parameter, where
superconductivity vanishes, is $g=g^{\pm}\approx$ 0.28 in the
$s^{\pm}$ pairing state. The extrapolated critical scattering
parameter, obtained using $\Delta\rho_{0}=$30 $\mu\Omega$ cm and
$\lambda=$260 nm, is $g^{\lambda}_{exp}\approx 3.7$. This value is
much larger than that expected in the $s^{\pm}$ scenario,
quantifying the break-down of the AG theory in irradiated
iron-arsenide superconductors, where the approximation of the
uniformly impurity-averaged Green's function is not valid.
Similar results were reported in
Ba(Fe$_{1-x}$Co$_{x}$)$_{2}$As$_{2}$ irradiated by
protons\cite{Kim} and illustrate the generality of the Swiss
cheese model for pair-breaking in this large class of
high-temperature superconductors.

\section{Conclusion}

We reported the influence of random point disorder
produced by  proton irradiation on the superfluid density in
Ca$_{0.5}$Na$_{0.5}$Fe$_2$As$_2$. It leads to a dramatic change of
$\lambda(0)$ after irradiation, in contrast to the small
variation of $T_{c}$ and predictions by the AG theory. Both
$\xi(T)$ and $\lambda(T)$ show similar temperature behavior
before and after irradiation. This behavior may be understood
within the Swiss cheese model, the pair-breaking nature of
$s^\pm$ interband superconductivity, and a short coherence
length, which considers the spatial dependence of the order
parameter and its strong suppression near defects at the atomic
scale. Finally, the extracted normalized scattering rate, in
conjunction with the absolute value of $\lambda(T)$, is much
larger than the critical scattering rate for the $s^{\pm}$
pairing, confirming the break-down of the AG theory in these
disordered superconductors. Further detailed multiband BdG model
calculations combined with systematic doping and irradiation
studies may shed light on the suppression of superconductivity in
this large class of iron-based superconductors.

Work at LANL was supported by the US Department of Energy, Basic
Energy Sciences, Division of Materials Sciences and Engineering.
Work at Brookhaven was supported by the US Department of Energy
under Contract No. DE-AC02-98CH10886. Work by G.F.C.
and W.Y. (fabrication of samples) was supported by the NSFC
under Grants No. 10974254 and No. 11074304, and by the
National Basic Research Program of China under Grants No.
2010CB923000 and No. 2011CBA00100. N.H. is member of CONICET
(Argentina).


\end{document}